# Comparison of conformal gravitation theories by solutions analogous to Reissner- Nordström ones

**M.V. Gorbatenko, S.Yu. Sedov**

Russian Federal Nuclear Center - All-Russia Research Institute of Experimental Physics, Sarov, Nizhny Novgorod Region

**Abstract**

An exact solution was produced for one of the versions of conformal -invariant gravitation theories (conformal geometrodynamics - CG) for the body with a mass and an electric charge. The solution is analogous to the Reissner-Nordström solution in the general relativity theory. Its similar solution was earlier produced by Mannheim and Kazanas for the different version of the conformal-invariant gravitation theory – Weyl conformal gravitation theory (Weyl CGT). Similar solutions do not comply. Issues of conformity with regard to the similar solutions are discussed. A conclusion is made that only the similar solution produced from conformal geometrodynamics (CG) equations conforms to electrodynamic equations.

# 1. INTRODUCTION

A number of exact solutions of the conformal gravitation theories have been found recently. So, in [1], an analogy of the external Schwarzschild solution known in general relativity theory (GR) was found. In [2], this solution was confirmed for another version of the conformal gravitation theory. Among the analogies of exact solutions of GR equations met in [3] there was produced an analogy of the Reissner-Nordström (RN) solution for the body with a mass and an electric charge.

It is clear that the found similar solutions are valuable themselves. But the situation with RN similar solution is new with respect to the fact that it is the solution of the conformal gravitation equations with a non-zero tensor of the matter. This fact happens to have far-reaching consequences; in particular, the produced RN similar solutions differ among themselves for different versions of the conformal gravitation theories.

Here, we shall consider two versions of the conformal gravitation theories different in the order of the derivatives in the equations:
- Weyl conformal gravitation theory (CGT) based on the equations with the derivatives from the forth-order metrics [4], [5].
- The theory which is a minimum conformal-invariant extension of GR equations and is based on the equations with the second-order derivatives from the metrics [6], [7]. The theory will be further named as conformal geometrodynamics (CG).

At the zero energy - momentum tensor any solution of the equation of one of the specified versions is an instantaneous solution of another version. This is testified also by the compliance of the similar solutions of the external Schwarzschild solution. However, when the energy - momentum tensor is different from zero, there may be no such compliance, that is also testified by the RN similar solutions. Providing the evidence for the given statement with regard to RN similar solutions is, in fact, one of the components of this work.

Before this proof we provide the detailed development of a RN similar solution for CG equations. As we know this result is a new one. Further the paper will compare RN similar solutions of the equations of two versions of the conformal gravitation theories from the point of view of their potential to reproduce self-consistent motion of particles with masses and electric charges and that are described with an electrodynamic equation consequent from the Darwin Langrangian (e.g., see §65 in [8]). We use the fact that CG equations in the region of zero Weyl vector concur with GR equations and reproduce all the attributes of electromagnetic interaction of particles with a mass and a charge. For GR equations it is a firmly established fact proved in [11], for example, using Einstein-Infeld-Hofmann method (EIH, [9], [10]).

The comparison allows us to make a conclusion that a similar solution of CG equations agrees with the experimental data, but a similar solution of Weyl CGT does not. Though this conclusion is preliminary, the way to include energy-momentum tensor accepted in Weyl CGT equations causes objections as it results into the solutions contradictive to the classical electrodynamics. It allows us to state that CG as a conformal gravitation theory has some advantages as compared to Weyl CGT.

## 2. Reissner-Nordström-Similar Solution For CG Equations

### 2.1. An external Schwarzschild-similar solution for CG equations

Central symmetric problem for CG equations without energy-momentum tensor of the matter, but with the account for lambda-term was solved in [2]. There was used a presumption that the metric is written in the form as

$$ds^2 = -e^{\gamma} dt^2 + e^{-\gamma} dr^2 + r^2 \left[ d\theta^2 + \sin^2\theta d\varphi^2 \right], \tag{1}$$

and the Weyl vector was in the form

$$A_\alpha = (0, A, 0, 0). \tag{2}$$

$A_1$ component of the Weyl vector $A_\alpha$ is marked just as $A$, as it comes from (2). CG equations in this case are brought to the following three equations:

$$\binom{0}{0} \Rightarrow e^{\gamma}\left[\frac{\gamma'}{r} + \frac{1}{r^2}\right] - \frac{1}{r^2} = e^{\gamma}\left[-A^2 - 2A' - \gamma' A - \frac{4}{r}A\right] + \lambda, \tag{3}$$

$$\binom{1}{1} \Rightarrow e^{\gamma}\left[\frac{\gamma'}{r} + \frac{1}{r^2}\right] - \frac{1}{r^2} = e^{\gamma}\left[-3A^2 - A\gamma' - \frac{4}{r}A\right] + \lambda, \tag{4}$$

$$\binom{2}{2} \Rightarrow e^{\gamma}\left[\frac{\gamma''}{2} + \frac{\gamma'^2}{2} + \frac{\gamma'}{r}\right] = e^{\gamma}\left[-A^2 - 2A' - 2\gamma' A - \frac{2}{r}A\right] + \lambda. \tag{5}$$

It was found out that the general exact solution of equations (3)-(5) has the view

$$\left. \begin{array}{c} ds^2 = -F(R) dt^2 + \dfrac{dR^2}{F(R)} + R^2 \left[ d\theta^2 + \sin^2\theta d\varphi^2 \right], \\[6pt] A = \dfrac{1}{(R_0 - R)}, \quad \lambda = \dfrac{\lambda_0 R_0^2}{(R_0 - R)^2}. \end{array} \right\} \tag{6}$$

Here, function $F(R)$ has the form of

$$F(R) = \left(1 - \frac{R}{R_0}\right)^2 \left(1 + \frac{r_0}{R_0} - \frac{r_0}{R}\right) + \frac{\lambda_0 R^2}{3}. \tag{7}$$

In (6), (7), quantities $r_0, \lambda_0$ are a gravitational radius and a cosmological constant, and $R_0$ is an integration constant. Radial variable $R$ is related to the initial radial alternative $r$ with the following ratio

$$R = \frac{r}{\left(1 + \dfrac{r}{R_0}\right)}, \quad r = \frac{R}{\left(1 - \dfrac{R}{R_0}\right)}. \tag{8}$$

In [2] solution (6), (7) was produced with conformal and coordinate transformations of the Schwarzschild solution in de Sitter space. The same solution was reproduced also by the way of the forth integration of CG equations (3)-(5).

With regard to the centrally symmetric problem Mannheim and Kazanas (MK) earlier in [1] found the solution of Weyl CGT equations in the following form:

$$ds^2 = -\left(e^{\gamma}\right)_{MK} dt^2 + \left(e^{-\gamma}\right)_{MK} dR^2 + R^2 \left[ d\theta^2 + \sin^2\theta d\varphi^2 \right], \tag{9}$$

where

$$\left(e^{\gamma}\right)_{MK} = (1-3\beta\gamma) - \frac{\beta(2-3\beta\gamma)}{R} + \gamma R - \kappa R^2 . \tag{10}$$

As it can be easily found, function $\left(e^{\gamma}\right)_{MK}$ in (10) is equivalent to function $F(R)$ in (7) with the precision up to the redefinition of constants:

$$\beta = \frac{r_0}{R_0\left(2+3\frac{r_0}{R_0}\right)}, \quad \gamma = -\frac{\left(2+3\frac{r_0}{R_0}\right)}{R_0}, \quad \kappa = -\frac{1}{R_0^2}\left(1+\frac{r_0}{R_0}\right) + \frac{\lambda_0}{3} \right\}. \tag{11}$$

So, Weyl CGT equations and CG equations result in the same square of the interval in the problem in case of Schwarzschild definition. In case Riemann properties of the space conform in both versions of the conformal gravitation theories the solution of CG equations(6), (7) has additional degrees of freedom in the form of Weyl vector and lambda-term that result into the $ds^2$ term linear by the radius.

### 2.2. REISSNER-NORDSTRÖM SOLUTIONS FOR GRT EQUATIONS IN DE SITTER SPACE

The algorithm when we find solutions of CG equations with conformal and coordinate transformations of the respective solution of the GR equations in de Sitter space used in [2] is evidently the simplest way to find such solutions. But for us to implement it we need to have a respective solution of GRT equations. Aiming at this, let's provide the Reissner-Nordström solution for the GR equations with asymptotics of the de Sitter space.

We base on the fact that in case $\lambda = 0$ the square of the interval that is written like

$$ds^2 = -\left[1 - \frac{r_0}{r} + \frac{Q^2}{r^2}\right]dt^2 + \frac{dr^2}{\left[1 - \frac{r_0}{r} + \frac{Q^2}{r^2}\right]} + r^2\left[d\theta^2 + \sin^2\theta d\varphi^2\right] \tag{12}$$

is a standard Reissner-Nordström solution. The metrics satisfies GR equations

$$G_{\alpha\beta} = \frac{8\pi G}{c^4} T_{\alpha\beta}^{(ph)} , \tag{13}$$

where $T_{\alpha\beta}^{(ph)}$ is the electromagnetic field tensor

$$T_{\alpha\beta}^{(ph)} = \frac{1}{4\pi}\left\{\Phi_{\alpha\sigma}\Phi_{\beta}^{\ \sigma} - \frac{1}{4}g_{\alpha\beta}\left(\Phi_{\mu\nu}\Phi^{\mu\nu}\right)\right\} . \tag{14}$$

Here, using $\Phi_{\alpha\beta}$ we identify tensor

$$\Phi_{\alpha\beta} = \Phi_{\beta,\alpha} - \Phi_{\alpha,\beta} , \tag{15}$$

where $\Phi_{\alpha}$ is a 4-vector- potential of the electromagnetic field (we keep symbol $A_{\alpha}$ reserved for the Weyl vector). That tensor $T_{\alpha\beta}$, which is a part of the dynamic equation, is related to tensor (14) with ratio

$$T_{\alpha\beta} = \frac{8\pi G}{c^4} T_{\alpha\beta}^{(ph)} . \tag{16}$$

Instead of electric charge $e$ it is convenient to introduce $Q$ constant with the length

$$Q \equiv \frac{\sqrt{G}\, e}{c^2}. \tag{17}$$

In case of the centrally symmetric problem, the component of $\Phi_{\alpha\beta}$ tensor different from zero will be

$$\Phi_{01} = -\frac{e}{r^2} = -\frac{c^2 Q}{G r^2}. \tag{18}$$

With the account of (18) we find that components of $T_{\alpha\beta}$ tensor different from zero are equal to:

$$T_{00} = e^\gamma \frac{Q^2}{r^4}, \tag{19}$$

$$T_{11} = -e^{-\gamma} \frac{Q^2}{r^4}, \tag{20}$$

$$T_{22} = \frac{Q^2}{r^2}. \tag{21}$$

GR equations (13) in case of the centrally symmetric problem are written in the form of

$$\binom{0}{0} \Rightarrow e^\gamma \left[\frac{\gamma'}{r} + \frac{1}{r^2}\right] - \frac{1}{r^2} = -\frac{Q^2}{r^4}, \tag{22}$$

$$\binom{1}{1} \Rightarrow e^\gamma \left[\frac{\gamma'}{r} + \frac{1}{r^2}\right] - \frac{1}{r^2} = -\frac{Q^2}{r^4}, \tag{23}$$

$$\binom{2}{2} \Rightarrow e^\gamma \left[\frac{\gamma''}{2} + \frac{\gamma'^2}{2} + \frac{\gamma'}{r}\right] = \frac{Q^2}{r^4}. \tag{24}$$

RN–similar solution in case of asymptotics of the de Sitter space exists and has the following square of the interval

$$ds^2 = -\left[1 - \frac{r_0}{r} + \frac{Q^2}{r^2} + \frac{\lambda_0}{3} r^2\right] dt^2 + \frac{dr^2}{\left[1 - \frac{r_0}{r} + \frac{Q^2}{r^2} + \frac{\lambda_0}{3} r^2\right]} + r^2 \left[d\theta^2 + \sin^2\theta d\varphi^2\right]. \tag{25}$$

Instead of (19)-(21), energy-momentum tensor components have the view

$$T_{00} = e^\gamma \left\{\frac{Q^2}{r^4} - \lambda_0\right\}, \tag{26}$$

$$T_{11} = -e^{-\gamma} \left\{\frac{Q^2}{r^4} - \lambda_0\right\}, \tag{27}$$

$$T_{22} = r^2 \left\{\frac{Q^2}{r^4} + \lambda_0\right\}. \tag{28}$$

And GRT equations instead of (22)-(24) are written as follows

$$\binom{0}{0} \Rightarrow e^\gamma \left[\frac{\gamma'}{r} + \frac{1}{r^2}\right] - \frac{1}{r^2} = -\frac{Q^2}{r^4} + \lambda, \tag{29}$$

$$\binom{1}{1} \Rightarrow e^\gamma \left[\frac{\gamma'}{r} + \frac{1}{r^2}\right] - \frac{1}{r^2} = -\frac{Q^2}{r^4} + \lambda, \tag{30}$$

$$\binom{2}{2} \Rightarrow e^\gamma \left[ \frac{\gamma''}{2} + \frac{\gamma'^2}{2} + \frac{\gamma'}{r} \right] = \frac{Q^2}{r^4} + \lambda . \tag{31}$$

Using direct substitution it is easy to confirm the correctness of formulas (25)-(31). We will not show it here.

### 2.3. REISSNER-NORDSTRÖM–SIMILAR SOLUTION FOR CG EQUATIONS

Let's find a solution similar to the Reissner-Nordström one for the CG equations. In case of the centrally symmetric problem CG equations with the account for non-zero energy-momentum tensor will have the following view instead of equations (3)-(5):

$$\binom{0}{0} \Rightarrow e^\gamma \left[ \frac{\gamma'}{R} + \frac{1}{R^2} \right] - \frac{1}{R^2} = -\frac{Q^2}{R^4} + e^\gamma \left[ -A^2 - 2A' - \gamma'A - \frac{4}{R}A \right] + \lambda , \tag{32}$$

$$\binom{1}{1} \Rightarrow e^\gamma \left[ \frac{\gamma'}{R} + \frac{1}{R^2} \right] - \frac{1}{R^2} = -\frac{Q^2}{R^4} + e^\gamma \left[ -3A^2 - A\gamma' - \frac{4}{R}A \right] + \lambda , \tag{33}$$

$$\binom{2}{2} \Rightarrow e^\gamma \left[ \frac{\gamma''}{2} + \frac{\gamma'^2}{2} + \frac{\gamma'}{R} \right] = \frac{Q^2}{R^4} + e^\gamma \left[ -A^2 - 2A' - 2\gamma'A - \frac{2}{R}A \right] + \lambda . \tag{34}$$

The radial variable in (32)-(34) is shown as $R$, the prime marks differentiation by $R$.

If we set quantities $A$ and $\lambda$ equal to zero in equations (32)-(34), then the solution will be a standard RN solution (12). If we set only quantity $A$ equal to zero, then we will get RN solution (25) with asymptotics of de Sitter space.

Let's use the operation algorithm from [2] to find the solution of CG equations (32)-(34). Following this algorithm we perform conformal and coordinate transformations and get:

$$\left. \begin{array}{l} ds^2 = -F(R)dt^2 + \dfrac{dR^2}{F(R)} + R^2 \left[ d\theta^2 + \sin^2\theta d\varphi^2 \right], \\[6pt] A' = \dfrac{1}{(R_0 - R)}, \quad \lambda = \dfrac{\lambda_0 R_0^2}{(R_0 - R)^2}. \end{array} \right\} \tag{35}$$

Here, function $F(R)$ has the form

$$F(R) = \left(1 - \frac{R}{R_0}\right)^2 \left(1 + \frac{r_0}{R_0} - \frac{r_0}{R} + \frac{Q^2}{R_0^2}\left(1 - \frac{R}{R_0}\right)^2\right) + \frac{\lambda_0 R^2}{3} . \tag{36}$$

In (6), (7) quantities $r_0, \lambda_0, Q$ are a gravitational radius, a cosmological constant and an electric charge, $R_0$ is an integration constant. Radial variable $R$ is related to the initial radial variable $r$ with ratio (8). After replacement of $R$ with $r$ we get the final expression for function $F(R)$ with the help of ratio (8):

$$F(R) = \left(1 + 3\frac{r_0}{R_0} + 6\frac{Q^2}{R_0^2}\right) - \left(r_0 + 4\frac{Q^2}{R_0}\right)\frac{1}{R} + \frac{Q^2}{R^2} - \\ - \left(\frac{2}{R_0} + 3\frac{r_0}{R_0^2} + 4\frac{Q^2}{R_0^3}\right)R + \left(\frac{1}{R_0^2} + \frac{r_0}{R_0^3} + \frac{Q^2}{R_0^4} + \frac{\lambda_0}{3}\right)R^2 . \tag{37}$$

Formulas (35), (37) are the similar solution for the RN one for CG equations. If in (35), (37) we set $Q = 0$, we shall get an MK solution [3].

With regard to the RN solution provided in [3] an interesting question arises: within standard GRT, what right part does metric (1) with the following function correspond to?

$$e^{\gamma} = w + \frac{u}{r} + v \cdot r - \kappa \cdot r^2 ? \tag{38}$$

That is with that function $e^{\gamma}$, which, according to [3], is a component of metric $-g_{00}$ in case when Weyl equations of CGT are solved with non-zero tensor of the electromagnetic field. Computations of the left parts of these equations by formulas (3)-(5) produce:

$$G_0^0 = \frac{(w-1)}{r^2} + \frac{2v}{r} - 3\kappa; \quad G_1^1 = G_0^0; \quad G_2^2 = \frac{v}{r} - 3\kappa. \tag{39}$$

Components $G_0^0, G_1^1, G_2^2$ within standard GR should have the sense of the components of the energy-momentum tensor of the electromagnetic field $T_0^0, T_1^1, T_2^2$. As quantities $G_0^0, G_1^1, G_2^2$ (39) do not have terms proportional to $\sim 1/r^4$, then quantities $T_0^0, T_1^1, T_2^2$ within a standard GR can't be components of the energy-momentum tensor of the electromagnetic field.

The difference in interpretation of the right parts in GR equations and interpretation of the Weyl CGT established above evidently can be considered as a test for selection of that version of the conformal generalization of GRT equations that agrees with experimental data.

## 3. RN – Similar Solution For CGT Weyl Equations

### 3.1. What is a Weyl conformal gravitation theory

We will give a more detailed explanation what Weyl CGT means. Conformal gravitation in a general sense is a theory, where invariance of the equations is observed with regard to metric transformation $g_{\alpha\beta} \to \Omega^2(x) \cdot g_{\alpha\beta}(x)$. This invariance can be reached in different ways. Weyl manifold is the basis in CG. Another way is to use a Lagrangian, which is squared by Weyl tensor or Riemann tensor. In Weyl CGT [1], Riemannian manifold is the basis as space-time and the operation for gravitation field is given

$$S = -\alpha \int d^4 x \sqrt{-g} \, C_{\alpha\beta\gamma\delta} C^{\alpha\beta\gamma\delta}, \tag{40}$$

where Weyl tensor $C_{\alpha\beta\mu\nu}$ is defined using Riemann tensor

$$C_{\alpha\beta\mu\nu} = R_{\alpha\beta\mu\nu} - \frac{1}{2}\left(g_{\alpha\mu} R_{\beta\nu} + g_{\beta\nu} R_{\alpha\mu} - g_{\alpha\nu} R_{\beta\mu} - g_{\beta\mu} R_{\alpha\nu}\right) + \frac{1}{6} R\left(g_{\alpha\mu} g_{\beta\nu} - g_{\alpha\nu} g_{\beta\mu}\right). \tag{41}$$

Equations for gravitation field without matter are called Bach equations:

$$B_{\alpha\beta} = B_{\alpha\beta}^{(1)} + B_{\alpha\beta}^{(2)} = 0, \tag{42}$$

where

$$B_{\alpha\beta}^{(1)} = -R_{\alpha\beta}{}^{;\nu}{}_{;\nu} + R^{\nu}{}_{\alpha;\beta;\nu} + R^{\nu}{}_{\beta;\alpha;\nu} - \frac{2}{3} R_{;\alpha;\beta} + \frac{1}{6} g_{\alpha\beta} R^{;\nu}{}_{;\nu}, \tag{43}$$

$$B_{\alpha\beta}^{(2)} = \frac{2}{3} R R_{\alpha\beta} - 2 R_{\alpha\nu} R^{\nu}{}_{\beta} - \frac{1}{6} R^2 g_{\alpha\beta} + \frac{1}{2} g_{\alpha\beta} R_{\mu\nu} R^{\mu\nu}. \tag{44}$$

If we introduce conformal - invariant energy-momentum tensor $T_{\alpha\beta}$, then CGT equations have the form $4\alpha \cdot B_{\alpha\beta} = T_{\alpha\beta}$, where $\alpha$ is a dimensionless factor (Weyl CGT theory parameter).

The advantages of this approach lie in the fact that basing on vacuum central- symmetric solution of Bach equations that results into Mannheim – Kazanas metric (9) we manage to explain rotational curves of galaxies, which are related to the dark matter problem (e.g., see [12]).

Let us note that this approach has some drawbacks. There are some problems with the description of gravitation waves in Weyl CGT. It was noted in [17] that the amplitude of the gravitational wave in vacuum in case of Weyl CGT does not decrease with the growth of the distance from the object with a non-zero energy-momentum tensor $T_{\alpha\beta}$. Another problem is the presence of growing solutions for gravitational perturbations of metric $h_{\alpha\beta}$, which are produced when solving linearized Bach equations when choosing Lorentz gauge $\partial^\alpha h_{\alpha\beta} = 0$: $\Box\Box h_{\alpha\beta} = 0$. Here, $g_{\alpha\beta} = \eta_{\alpha\beta} + h_{\alpha\beta}$, $\eta_{\alpha\beta}$ - is metrics of pseudo-Euclidean space, $\Box$ –is D'Alembertian.

### 3.2. RN similar solution for Weyl CGT.

Paper [3] offers the following version of electromagnetic field introduction. It is known that in common space-time with dimension equal to 4 Maxwell equations are conformally invariant (which disappears in spaces of higher dimensionality). So, Maxwell equations and energy-momentum tensor of the electromagnetic field are compatible with Weyl CGT structure in 4-D space-time. Let a system of units is used with $c = 1$. The operation has the form:

$$S = -\int d^4 x \sqrt{-g} \left( \alpha \cdot C_{\alpha\beta\gamma\delta} C^{\alpha\beta\gamma\delta} - \frac{1}{4} \Phi_{\mu\nu}\Phi^{\mu\nu} - \Phi_\mu J_\mu \right), \qquad (45)$$

where $\Phi_{\mu\nu}$ is the electromagnetic field tensor, $\Phi_\nu$ is vector electromagnetic potential, $J_\mu$ is current vector, $\Phi_{\mu\nu} = \frac{\partial \Phi_\nu}{\partial x^\mu} - \frac{\partial \Phi_\mu}{\partial x^\nu}$.

For us to account for the electric $Q$ and magnetic $P$ charges in Schwarzschild metrics, the authors in [3] introduce vector potential of electromagnetic field:

$$\Phi_\mu = \left( \frac{Q}{r}, 0, 0, -P\cos\theta \right). \qquad (46)$$

This potential results into the quantity of the electric field with a non-zero component $E_r = F_{01} = \frac{Q}{r^2}$ and magnetic field $H_r = \frac{1}{2}\varepsilon_{01\sigma\tau}F^{\sigma\tau} = \frac{P}{r^2}$ [3]. Maxwell equations of $\Phi^{\mu\nu}{}_{;\nu} = 0$ form are fulfilled in the region free from charges. Electromagnetic energy-momentum tensor is identified as

$$T^{(ph)\alpha\beta} = -\Phi^{\alpha\gamma}\Phi^\beta_\gamma + \frac{1}{4} g^{\alpha\beta}\Phi_{\gamma\delta}\Phi^{\gamma\delta}.$$

So, Weyl CGT equations are satisfied in the area without particles of the matter

$$4\alpha \cdot B_{\alpha\beta} = T^{(ph)}{}_{\alpha\beta} \qquad (47)$$

with conformally invariant electromagnetic energy-momentum tensor that has radial component $T^{(ph)r}{}_r = -\left(Q^2 + P^2\right)/2r^4$ [3].

The square of metrics for this CGT problem is written as [3]

$$ds^2 = -B(r)dt^2 + dr^2/B(r) + r^2 d\Omega, \qquad (48)$$

where $r$ is radial variable, $d\Omega$ is as part of solid angle,

$$B(r) = w + \frac{u}{r} + v \cdot r - \kappa \cdot r^2 = 1 - 3\beta\gamma - \frac{\beta(2-3\beta\gamma) + \frac{Q^2+P^2}{8\alpha}\gamma}{r} + \gamma r - kr^2, \qquad (49)$$

$\beta$, $\gamma$ and $k$ are free parameters [3]. Quantity $\beta$ is directly related to the value of the point mass $M$ located in the center of the coordinates, quantity $k$ is interpreted as a cosmological constant, and quantity $\gamma$ is determined from the rotational curves of galaxies rotation. Let's note that for GRT (see [8]) the square of metrics in the field of the point mass is set at $G=1$, $c=1$ with ratios:

$$ds^2 = -B(r)dt^2 + dr^2/B(r) + r^2 d\Omega, \quad B(r) = \left(1 - \frac{2M}{r} + \frac{Q^2}{r^2}\right), \qquad (50)$$

i.e. the dependence of the charge part in the metrics on the radial coordinate in GRT is quite different; it is $\sim \frac{1}{r^2}$.

### 3.3 Transfer to Newtonian limit in Weyl CGT

The equation for trial uncharged mass $m$ for GR in Newtonian limit in a static case has the following form [8]:

$$ds^2 \approx -(1 + 2\frac{\varphi}{c^2}) \cdot dt^2 + dr^2 + r^2 d\Omega, \quad m\ddot{\vec{r}} = m \cdot \nabla\varphi, \qquad (51)$$

where $\varphi = -\frac{GM}{r} + \frac{Q^2}{2c^2 r^2}$. In the region without a source at $r > r_0$ then $\nabla^2\varphi = 0$. Let velocity of light will further be $c = 1$.

In the case of Weyl CGT in Newtonian limit

$$ds^2 \approx -(1 + 2\varphi) \cdot dt^2 + dr^2 + r^2 d\Omega, \quad m\ddot{\vec{r}} = m \cdot \nabla\varphi, \qquad (52)$$

where $\varphi \approx -\frac{\beta(2-3\beta\gamma) + Q^2\gamma/8\alpha}{2r} + \frac{\gamma}{2}r$. Let's identify source function $f$ related to the baryon masses (neutrons and protons) as $f(r) = \frac{3}{4\alpha B(r)}(T_0^0 - T_r^r)$ [1]. In Newtonian limit when the mass is non-zero and with no account for electromagnetic field $f(r) \approx \frac{3}{4\alpha}T_0^0$. Quantity $\varphi$ in Weyl CGT is the solution of the equation with source $\nabla^4\varphi = \frac{1}{2}f(r)$ [17]. Let's presume $f(r) = 0$ at $r > r_0$ (assuming that the mass is concentrated in the region $r \leq r_0$), then (e.g., see [17]):

$$\varphi(r) = -\frac{1}{3r}\int_0^{r_0} dr' r'^4 f(r') - r\int_0^{r_0} dr' r'^2 f(r'). \qquad (53)$$

For the case of one point mass $M$ in the center $r_0 \to 0$.

This implies that source function $f$ in Weyl CGT at the presence of charge $Q$ and mass $M$ has imposed conditions:

$$-\frac{1}{3}\int_0^{r_0} dr' r'^4 f(r') = -\frac{\beta(2-3\beta\gamma) + Q^2\gamma/8\alpha}{2}, \quad \int_0^{r_0} dr' r'^2 f(r') = -\frac{\gamma}{2}. \quad (54)$$

Let's take charge $Q = 0$. If we admit that quantity $\beta(2-3\beta\gamma)$ is equal to mass $M$, then quantities $\beta$ and $\gamma$ depend on the particular kind of distribution of the continuous source function $f(r)$ (as some analogy of the regular density of the matter $\rho$) inside radius $r \leq r_0$. It was noted in [17], for example. To get rid of this unpleasant property Mannheim takes that source function $f(r)$ as discontinuous (e.g., see [18]). It is determined by the contribution of point massive particles of number $N$ within radius $r < r_0$ (baryon) [18]:

$$\frac{1}{2}f(r<r_0) \equiv -\gamma \sum_{i=1}^{N}\frac{\delta(\vec{r}-\vec{r}_i)}{r^2} - \frac{3\beta(2-3\beta\gamma)}{2}\sum_{i=1}^{N}\left[\nabla^2 - \frac{r^2}{12}\nabla^4\right]\frac{\delta(\vec{r}-\vec{r}_i)}{r^2}. \quad (55)$$

In case charge $Q$ is present it is worth presuming

$$\frac{1}{2}f(r<r_0) = -\gamma \sum_{i=1}^{N}\frac{\delta(\vec{r}-\vec{r}_i)}{r^2} - \frac{3\beta(2-3\beta\gamma)+3Q^2\gamma/8\alpha}{2}\sum_{i=1}^{N}\left[\nabla^2 - \frac{r^2}{12}\nabla^4\right]\frac{\delta(\vec{r}-\vec{r}_i)}{r^2}. \quad (56)$$

With values $\gamma \approx 0$ and $Q = 0$ the results correspond to the Newtonian limit $\varphi = -\frac{GNM}{r}$ at the radius of $r > r_0$ and the choice of function $\beta = GNM$ for $N$ of closely located particles. But at $Q \neq 0$ addition of squared charges of $N$ point particles (baryons) takes place in function $f(r)$. With regard to this the following fact is revealed in the Weyl CGT. It could seem that the presence of $N$ closely located compensated point charges of the opposite sign $\pm Q$ (if there are $N/2$ each of the particles of a definite sign) should result into the zero contribution of the full charge of the neutral liquid into the similar potential of the gravitational potential $\varphi$ in Newtonian approximation. But in Weyl CGT theory it is not so; the contribution of the charged part into potential $\varphi$ appears to be proportional to $NQ^2$ because of discontinuity of $f$ and it increases with the growth of $N$, as well as mass $NM$. That is in fact in Weyl CGT in the presence of charge $e$ gravitational mass of the proton is renormed from quantity $m_p$ to quantity $m_p + \frac{e^2\gamma}{16c^2\alpha}$. This is acceptable only if inequality $\frac{e^2\gamma}{16\alpha} << m_p c^2$ is valid (value $\gamma^{-1}$ is equal to some characteristic length of Weyl CGT). So, in Weyl CGT gravitational and electromagnetic effects are "fully mixed".

So, we see that there are particular problems in Weyl CGT either with the experiments of Cavendish type in case of continuous mass distribution or with realization of the weak equivalence principle if the baryon particles have a charge.

## 4. DISCUSSION

By now we have two similar solutions for the one of Reissner-Nordström:

- The solution, which is provided in [3] called as «Reissner-Nordström solution for Weyl CGT equations». It has the form given in (48), (49).
- A similar solution of the RN solution (35)-(37) for CG equations.

The two specified solutions similar to the RN one evidently do not coincide. There is a question about what physical consequences result from both solutions.

Let's do the replacement of the metrics of each RN similar solution in the left part of the expression for Einstein tensor $G_\alpha^\beta$. Here, components $G_0^0, G_1^1, G_2^2$ should be equal to the right parts of expressions (29)-(31). In case of CG this condition is fulfilled, but in case of Weyl CGT all listed components $G_0^0, G_1^1, G_2^2$ are equal to lambda-term.

Let's try to produce an electrodynamic equation that follows from Darwin Lagrangian and describes self-consistent motion of particles with a mass and a charge. This problem was considered in a number of works (for example, in [13]-[15], [9]). In [13], [15], [9] the problem was solved using Einstein-Infeld-Hoffmann method, and in [14] it was solved with Fock method. In our opinion, the most consistent results are in [9], where Einstein-Infeld-Hoffmann method reproduced the equation that matches up Darwin Lagrangian for the energy-momentum tensor described with formulas (19)-(21). Only RN-similar solution for CG equations can bring to this result, as in case of CG, energy-momentum tensor complies with expressions (19)-(21) (with the precision to lambda-term of cosmological origin). Component $g_{00}$ in the Weyl CGT solution ((48),(49)) does not contain terms $\sim 1/r^2$ and so it can't reproduce electrodynamic effects.

In case of Weyl CGT, the problem of self-consistent motion of particles with a mass and a charge can also be solved. But in this case some forces analogous to Newtonian and post-Newtonian ones are produced with effective parameters of particles that are combinations of a mass and a charge. A specific feature of the particles motion is no strict law for charge conservation and the presence of electro-gravitational effects. Evidently, this makes Weyl CGT not enough adequate to the reality. Besides this problem, there are some other issues in Mannheim theory [17].

As it was noted by the authors in [16], if the energy-momentum tensor of the matter is equal to zero, then any solution of CG equations is automatically a solution of Weyl CGT equations. But if the energy-momentum tensor of the matter is not equal to zero, CG solutions may be not the solutions of Weyl CGT, and vice versa. An RN solution provided in [3] and the solution (35) produced in this work with function $F(R)$ in the form of (36), or (37) confirm this directly.

We suppose that two factors played a key role here. First of all, it was the fact that the energy-momentum tensor of the matter is not equal to zero. Secondly, it was the fact that this tensor in case of Weyl CGT is the right part for the equations with the forth derivatives, and in case of CG it is for the equations with the second derivatives.